\documentclass[iop,revtex4]{emulateapj} 

\usepackage{hyperref,enumerate,numdef,color}
\hypersetup{
    unicode=false,                  
    pdftoolbar=true,                
    pdfmenubar=true,                
    pdffitwindow=true,              
    pdfstartview={FitH},            
    pdftitle={G64-12 and G64-37},   
    pdfauthor={vplacco},            
    pdfsubject={Astronomy},         
    pdfcreator={dvipdf},            
    pdfproducer={dvipdf},           
    pdfkeywords={metal-poor stars}, 
    pdfnewwindow=true,              
    colorlinks=true,                
    linkcolor=red,                  
    citecolor=blue,                 
    filecolor=magenta,              
    urlcolor=cyan,                  
    breaklinks=true,
    linktocpage
}

\num\newcommand{\g12}{\object{G64$-$12}}
\num\newcommand{\g37}{\object{G64$-$37}}

\newcommand{\eps}[1]{\ensuremath{\log\epsilon\,(\mathrm{#1})}}

\newcommand{\abund}[2]{\ensuremath{[\mathrm{#1}/\mathrm{#2}]}}
\newcommand{\cfe}{\abund{C}{Fe}}

\newcommand{\xfe}[1]{\abund{#1}{Fe}}
\newcommand{\xh}[1]{\abund{#1}{H}}
\newcommand{\metal}{\abund{Fe}{H}}

\newcommand{\teff}{\ensuremath{T_\mathrm{eff}}}
\newcommand{\logg}{\ensuremath{\log\,g}}
\newcommand{\Msun}{\mathrm{M}_\odot}

\newcommand{\erg}{\ensuremath{\mathrm{erg}}}

\shorttitle{\g12\ and \g37\ are CEMP-no Stars}
\shortauthors{Placco et al.}

\begin{document}

\title{\g12\ and \g37\ are Carbon-Enhanced Metal-Poor Stars}

\author{
Vinicius M.\ Placco\altaffilmark{1,2},
Timothy   C.\ Beers\altaffilmark{1,2},\\
Henrique   Reggiani\altaffilmark{3},
Jorge    Mel\'endez\altaffilmark{3}}

\altaffiltext{1}{Department of Physics, University of Notre Dame, 
                 Notre Dame, IN 46556, USA}
\altaffiltext{2}{JINA Center for the Evolution of the Elements, USA}
\altaffiltext{3}{Departamento de Astronomia - Instituto de Astronomia, 
                 Geof\'isica e Ci\^encias Atmosf\'ericas, Universidade 
				 de S\~ao Paulo, S\~ao Paulo, SP 05508-900, Brazil}

\addtocounter{footnote}{3}

\begin{abstract}

We present new high-resolution chemical-abundance analyses for the
well-known high proper-motion subdwarfs \g12\ and
\g37, based on very high signal-to-noise spectra ($S/N \sim 700/1$) with
resolving power $R \sim 95,000$. These high-quality data enable the first
{\it reliable} determination of the carbon abundances for these two
stars; we classify them as carbon-enhanced metal-poor (CEMP) stars based
on their carbonicities, which both exceed [C/Fe] $= +1.0$.  They
are sub-classified as CEMP-no Group-II stars, based on their
location in the Yoon-Beers diagram of absolute carbon abundance, $A$(C)
vs. [Fe/H], as well as on the conventional diagnostic [Ba/Fe]. The
relatively low absolute carbon abundances of CEMP-no stars, in
combination with the high effective temperatures of these two stars
(\teff\ $\sim 6500$~K) weakens their CH molecular features to the point
that accurate carbon abundances can only be estimated from spectra with
very high $S/N$. A comparison of the observed abundance patterns with
the predicted yields from massive, metal-free supernova models 
reduces the inferred progenitor masses by factors of $\sim$ 2-3,
and explosion energies by factors of $\sim$ 10-15, compared to those
derived using previously claimed carbon abundance estimates. There are
certainly many more warm CEMP-no stars near the halo main-sequence
turnoff that have been overlooked in past studies, directly impacting
the derived frequencies of CEMP-no stars as a function of metallicity, a
probe that provides important constraints on Galactic chemical evolution
models, the initial mass function in the early Universe, and first-star
nucleosynthesis.

\end{abstract}


\section{Introduction}
\label{intro}

The basic cosmological framework suggesting that the chemical evolution
of the Universe is a continuous process of nucleosynthesis and mixing of
subsequent stellar generations has been in place since the work of
\citet{hoyle1954}. In this scenario, the long-lived, low-mass extremely
metal-poor \citep[EMP; \metal\footnote{\abund{A}{B} = $log(N_X/{}N_Y)
_{\star} - \log(N_X/{}N_Y) _{\odot}$, where $N$ is the number density of
atoms of elements $X$ and $Y$ in the star ($\star$) and the Sun
($\odot$), respectively.}~$<-3.0$, e.g.,][]{beers2005,frebel2015} stars
observed today were formed from gas that was polluted by the
nucleosynthesis products of previous generations of (likely massive)
stars, also known as Population III (Pop III). The proposed existence of
massive Pop III stars is not new \citep[e.g., ][]{puget1980}; the first
comparison between the theoretical yields of Pop III stars and observed
elemental abundances of very metal-poor stars was made by
\citet{nomoto1999}. Since then, there has been significant advances in our
understanding of the underlying physics of first-generation stars, and
different classes of models are able to well-reproduce observations of
EMP stars in the Galaxy \citep[see][for a brief summary]{placco2016}.

Despite the fact that the elemental-yield predictions of first-star
nucleosynthesis differ somewhat between various authors, they all agree
that the light element carbon plays a central role at early times.
Indeed, the majority of the long-lived relics 
of the chemical evolution in the early Universe are expected to be 
heavy-metal deficient, but carbon enhanced \citep{frebel2007b}. Such
stars, which once were considered an {\it{Astrophysical Enigma}} by
\citet{bidelman1956}\footnote{The author states ``{\it{The spectra of
these objects show extremely strong absorption features due to CH, and
considerably weaker lines of neutral metals than do the typical carbon
stars.}}''}, today are known as carbon-enhanced metal-poor (CEMP)
stars, following recognition of their existence at the lowest
metallicities by \citet{beers1992}.

Although carbon-enhanced Solar-metallicity stars were first identified
almost 75 years ago \citep{keenan1942}, the majority of early spectral
catalogs contained only cooler (\teff~$< 4500$~K) carbon stars with
strong molecular features \citep{kamijo1959}. One of the first attempts
to determine carbon abundances for warmer (\teff$> 5500$~K),
low-metallicity stars came some twenty years later \citep{peterson1978}.
This study only considered stars of relatively higher metallicity
(\metal~$>-2.3$), since few stars were known below this abundance at the
time. Nevertheless, it showed that, even when the carbon abundance ratio
relative to iron \citep[carbonicity;][]{placco2011} of a given star is
high (\cfe~$>+1.0$), the CH molecular features around 4300\,{\AA}\ can
be quite weak at these temperatures, present at less than 2\% of the
continuum level for \teff\ = 6500~K. As a result, reliable carbonicity
determinations for warm stars are biased towards relatively bright
stars, where the $S/N$ required to detect such small deviations from the
continuum can be obtained with reasonable integration times.

\begin{deluxetable*}{lrrrrrrrrrr}[!ht]
\tablewidth{40pc}  
\tabletypesize{\small}
\tablecaption{Final Abundance Estimates for \protect\g12\ and \protect\g37. \label{abfinal}} 
\tablehead{ 
&& \multicolumn{4}{c}{\g12} &  & \multicolumn{4}{c}{\g37} \\
 \cline{3-6} \cline{8-11}
Ion & $\log\epsilon_{\odot}$\,(X)  & $\log\epsilon$\,(X) 
                                   & $\mbox{[X/H]}$ & $\mbox{[X/Fe]}$ & $\sigma$ &
                                   & $\log\epsilon$\,(X) 
								   & $\mbox{[X/H]}$ & $\mbox{[X/Fe]}$ & $\sigma$  }
\startdata 
\ion{Li}{1} &  1.05 &    2.36 & $+$1.31 & $+$4.59 &  0.04 &&    2.25 & $+$1.20 & $+$4.31 &  0.03 \\
C (CH)      &  8.43 &    6.21 & $-$2.22 & $+$1.07 &  0.05 &&    6.44 & $-$1.99 & $+$1.12 &  0.05 \\
\ion{O }{1} &  8.69 &    6.58 & $-$2.11 & $+$1.17 &  0.07 &&    6.59 & $-$2.10 & $+$1.00 &  0.03 \\
\ion{Na}{1} &  6.24 &    2.87 & $-$3.37 & $-$0.09 &  0.03 &&    2.92 & $-$3.32 & $-$0.21 &  0.03 \\
\ion{Mg}{1} &  7.60 &    4.79 & $-$2.80 & $+$0.48 &  0.07 &&    4.87 & $-$2.73 & $+$0.38 &  0.07 \\
\ion{Al}{1} &  6.45 &    2.56 & $-$3.90 & $-$0.61 &  0.05 &&    2.63 & $-$3.82 & $-$0.71 &  0.05 \\
\ion{Si}{1} &  7.51 &    4.06 & $-$3.45 & $-$0.16 &  0.04 &&    4.11 & $-$3.40 & $-$0.29 &  0.03 \\
\ion{Ca}{1} &  6.34 &    3.56 & $-$2.78 & $+$0.50 &  0.03 &&    3.64 & $-$2.70 & $+$0.41 &  0.03 \\
\ion{Sc}{2} &  3.15 &    0.03 & $-$3.12 & $+$0.17 &  0.06 &&    0.20 & $-$2.95 & $+$0.15 &  0.05 \\
\ion{Ti}{1} &  4.95 &    2.40 & $-$2.55 & $+$0.74 &  0.05 &&    2.53 & $-$2.42 & $+$0.69 &  0.04 \\
\ion{Ti}{2} &  4.95 &    2.19 & $-$2.76 & $+$0.52 &  0.09 &&    2.34 & $-$2.61 & $+$0.50 &  0.07 \\
\ion{Cr}{1} &  5.64 &    2.27 & $-$3.37 & $-$0.09 &  0.05 &&    2.50 & $-$3.14 & $-$0.03 &  0.04 \\
\ion{Mn}{1} &  5.43 &    1.50 & $-$3.93 & $-$0.65 &  0.05 &&    1.77 & $-$3.66 & $-$0.55 &  0.04 \\
\ion{Fe}{1} &  7.50 &    4.21 & $-$3.29 &    0.00 &  0.04 &&    4.39 & $-$3.11 &    0.00 &  0.03 \\
\ion{Fe}{2} &  7.50 &    4.26 & $-$3.24 & $+$0.05 &  0.06 &&    4.45 & $-$3.05 & $+$0.05 &  0.04 \\
\ion{Co}{1} &  4.99 &    2.10 & $-$2.89 & $+$0.40 &  0.05 &&    2.23 & $-$2.76 & $+$0.35 &  0.05 \\
\ion{Ni}{1} &  6.22 &    2.95 & $-$3.27 & $+$0.02 &  0.03 &&    3.17 & $-$3.05 & $+$0.06 &  0.02 \\
\ion{Zn}{1} &  4.56 &    1.80 & $-$2.76 & $+$0.52 &  0.03 &&    1.92 & $-$2.64 & $+$0.47 &  0.02 \\
\ion{Sr}{2} &  2.87 & $-$0.34 & $-$3.21 & $+$0.07 &  0.06 && $-$0.19 & $-$3.06 & $+$0.05 &  0.04 \\
\ion{Ba}{2} &  2.18 & $-$1.17 & $-$3.35 & $-$0.07 &  0.06 && $-$1.28 & $-$3.46 & $-$0.36 &  0.04
\enddata
\end{deluxetable*}

The recognition that carbon-enhanced stars occur with
higher frequency at lower metallicities extends back over
two decades \citep[e.g.,][and references therein]{beers1992}. 
Subsequent observations of metal-poor stars using medium-
and high-resolution spectroscopy has confirmed that the fractions of
CEMP stars increases from $\sim 15-20$\% for \metal~$ < -2.0$ to more
than 80\% for \metal~$ < -4.0$ \citep{lee2013, placco2014c}, prima facia
observational evidence that carbon is an important contributor to the
chemical evolution of the early Universe.

Among the various sub-classes of CEMP stars \citep{beers2005}, the
CEMP-no stars (which exhibit no enhancements in their neutron-capture
elements, e.g., \xfe{Ba} $ < 0.0$) are believed to have formed from gas
polluted by the nucleosynthesis products of massive stars, perhaps
including Pop III progenitors. Even though the low abundances of
neutron-capture elements (such as Ba) is a feature of such stars, it has been
recently suggested by \citet{yoon2016}, building on the work of
\citet{spite2013} and \citet{bonifacio2015}, that the absolute carbon
abundance ($A$(C) $ = \log\,\epsilon $(C)) is a sufficient (and likely
more fundamental) criterion to distinguish CEMP-no stars from the far
more populous sub-class of CEMP-$s$ stars (where the carbon enhancement
arises due to mass transfer from a binary asymptotic giant-branch
stellar companion).  This new criterion is particularly useful for
warmer CEMP stars, where measurement of [Ba/Fe] can prove challenging. 

In this paper, we present the first confirmation that the well-studied
subdwarfs \g12\ and \g37\ are in fact CEMP-no stars. Even though these
stars have been considered by numerous previous authors, accurate
determinations of their carbon abundances was limited by the quality of
the available spectra. It is highly likely that other warm
low-metallicity stars have been incorrectly classified as
``carbon-normal,'' a deficiency that must be addressed in the future. 
This paper is outlined as follows: Section~\ref{secobs} describes the
spectroscopic data used in this work, and presents the
elemental-abundance estimates obtained for \g12\ and \g37\ from our very
high-quality data. Section~\ref{disc} considers the consequences of
properly classifying warm CEMP-no stars on our understanding the nature of
their progenitors, and on the derived CEMP fractions as a function of
metallicity. Our conclusions are provided in Section~\ref{final}.

\section{\protect\g12\ and \protect\g37: CEMP-no Stars}
\label{secobs}

\g12\ (Wolf~1492; $\alpha=13^h40^m02^s.498$,
$\delta=-00^{\circ}02'18.80''$, $V=11.45$) is a well-known, high
proper-motion EMP subdwarf. It was first identified as such by
\citet{sandage1969} and, since then, has been the target of 225 studies
in the literature, according to the SIMBAD database. It also possesses
an extremely high radial velocity \citep[$V_r = 442.51\pm0.18$~km/s;
][]{latham2002}, and indeed has been used to obtain early estimates of
the mass of the Galaxy from its derived rest-frame space velocity
\citep{carney1988}. Based on 33 radial-velocity
measurements spanning 4818 days, \citet{latham2002} concluded that \g12\
is not in a binary system. The full space motion of
\g12\ estimated by \citet{allen1991} clearly indicates that it is a
member of the outer-halo population \citep{carollo2007,
beers2012}; a high-energy retrograde orbit ($\Theta\ \sim -138$~km/s)
with a maximum distance from the Galactic plane $Z_{\rm max} \sim 40$~kpc. 

According to the SAGA database \citep{saga2008}, there are 26 papers with
estimated chemical abundances for \g12. However, the carbonicity for this star
is estimated by only four studies \citep{akerman2004, barklem2005,aoki2006,
zhang2011}, with values ranging from \cfe\ $= +0.30$ to \cfe\ $= +0.88$. Most
other studies either do not report an estimate of [C/Fe], or only obtain weak
upper limits. In addition, \citet{fabbian2009} derived the carbon abundance for
\g12\ using \ion{C}{1} features in the near-infrared, with a value of \cfe\
$= +0.53$.

\begin{figure*}[!ht]
\epsscale{1.15}
\plotone{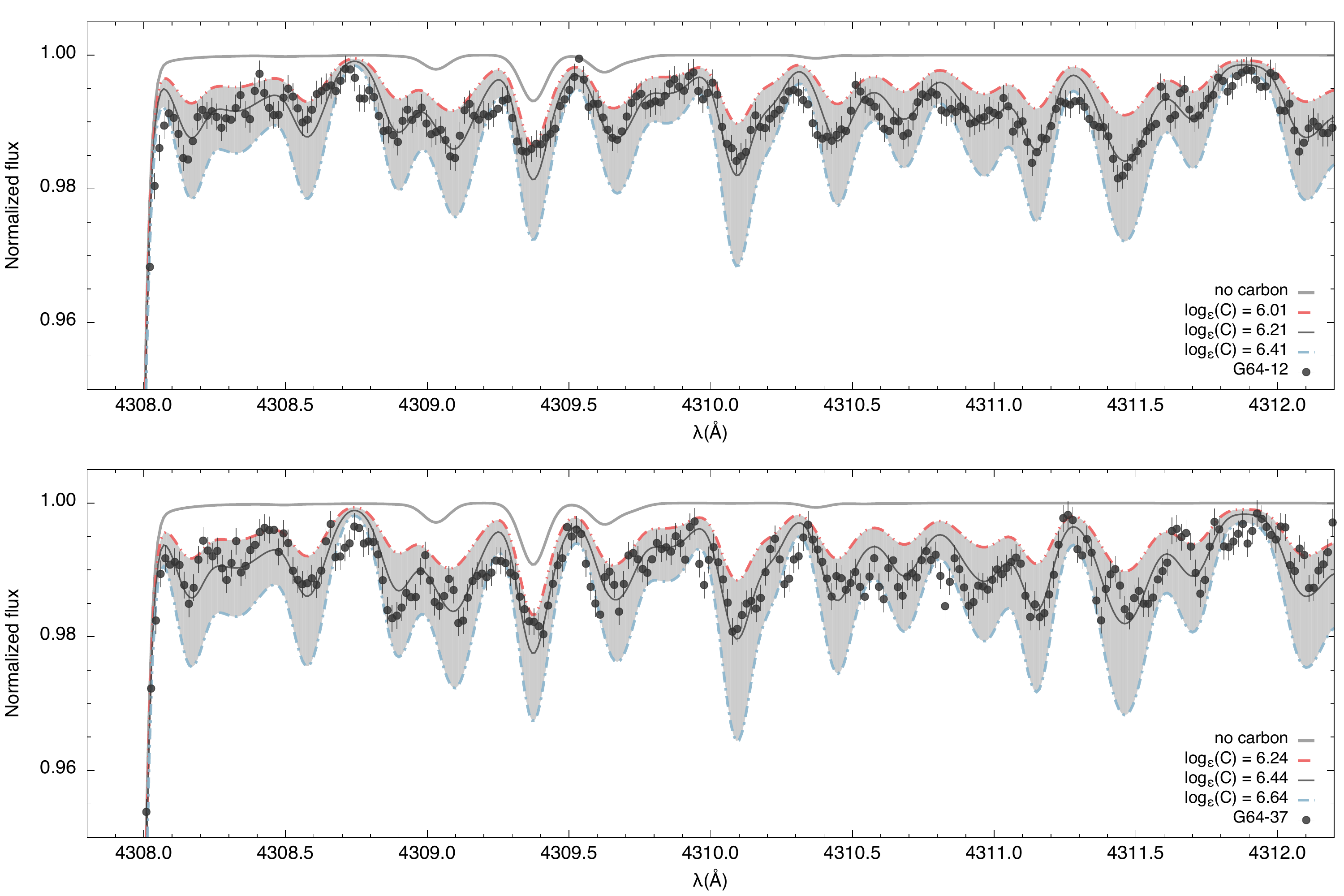}
\caption{Spectral synthesis of the CH $G$-band for \protect\g12\ (upper panel) and
\protect\g37\ (lower panel). The dots represent the observed spectra, the solid
black lines are the best abundance fits, and the dotted and dashed lines are the lower and
upper abundances, used to estimate the uncertainty. The shaded area
encompasses a 0.4~dex difference in \eps{C}. The light gray lines
show the synthesized spectrum in the absence of carbon.  Note the
extremely small maximum deviation of the carbon features for both stars
below the continuum, $\sim 1-2\%$.}
\label{csyn}
\end{figure*}

\g37\ (Ross~841; $\alpha=14^h02^m30^s.091$,
$\delta=-05^{\circ}39'05.20''$, $V=11.15$) is also an EMP subdwarf,
recognized as a high proper-motion star decades before \g12\
\citep{ross1930}. Although it is well-studied, this star has fewer
references in the literature than \g12, 115 according to the SIMBAD
database. Even though its radial velocity is lower \citep[$V_r =
81.52\pm0.20~$km/s;][]{latham2002}, its derived space motion also
indicates clear membership in the outer-halo population.
\g37\ is also considered a single star by
\citet{latham2002}, based on 22 radial velocity measurements spanning
1938 days. \citet{allen1991} obtain a high-energy retrograde orbit
($\Theta\ \sim -209$~km/s) with a maximum distance from the plane
$Z_{\rm max} \sim 10-35$~kpc (according to these authors, the large
errors in this estimate arise primarily from the sensitivity of its
derived orbit to distance errors). 

Elemental abundances for \g37\ are reported by 20 different studies,
according to the SAGA database, but measurements of carbonicity are reported
only by \citet{akerman2004}, \cfe\ $= +0.29$, and \citet{fabbian2009}, \cfe\
$= +0.41$.

The high-resolution spectra for \g12\ and \g37\ used in this work were
obtained with the HIRES instrument on the Keck 10-m telescope at Mauna
Kea. The resolving power of $R \sim 95,000$ and signal-to-noise ratio
$S/N = 700/1$ at 5000\,{\AA} allowed for an elemental-abundance analysis
of unprecedented precision for these two stars. The first high-quality
differential abundance analysis of EMP stars, based on \g12\ and \g37,
was reported by \citet{reggiani2016}; their derived 1D, LTE abundances
(which did not appear in the published work) are shown here for the
first time. The abundances were measured using the semi-automated code
{\texttt{q$^2$}} \citep{ramirez2014}, which communicates with the radiative
transfer code MOOG \citep{sneden1973}.

The atmospheric parameters adopted for these stars are: 
\g12\  [\teff = 6463(50)~K, \logg = 4.26(0.15), \metal = $-$3.29(0.02),
and $\xi$ = 1.65(0.06)~km/s];
\g37\  [\teff = 6570(27)~K, \logg = 4.40(0.06), \metal = $-$3.11(0.02), 
and $\xi$ = 1.74(0.06)~km/s]. 
The atmospheric parameters of \g37\ were calculated differentially, using the
Fe lines from \g12\ as a reference. The errors for \g12\ come from the
original sources of the atmospheric parameters (IRFM \teff\
from \citealt{melendez2010b} and \logg\ from \citealt{nissen2007},  and the 
differential calculations for \metal\ and $\xi$ from \citealt{reggiani2016}). 
The errors for \g37\ were calculated differentially by varying the atmospheric 
parameters of \g12.
Table~\ref{abfinal} lists the \eps{X}, \xh{X}, \xfe{X} abundances and
uncertainties for 17 elements. The Solar elemental abundances of 
\citet{asplund2009} were used.

\begin{figure*}[!ht]
\epsscale{1.15}
\plotone{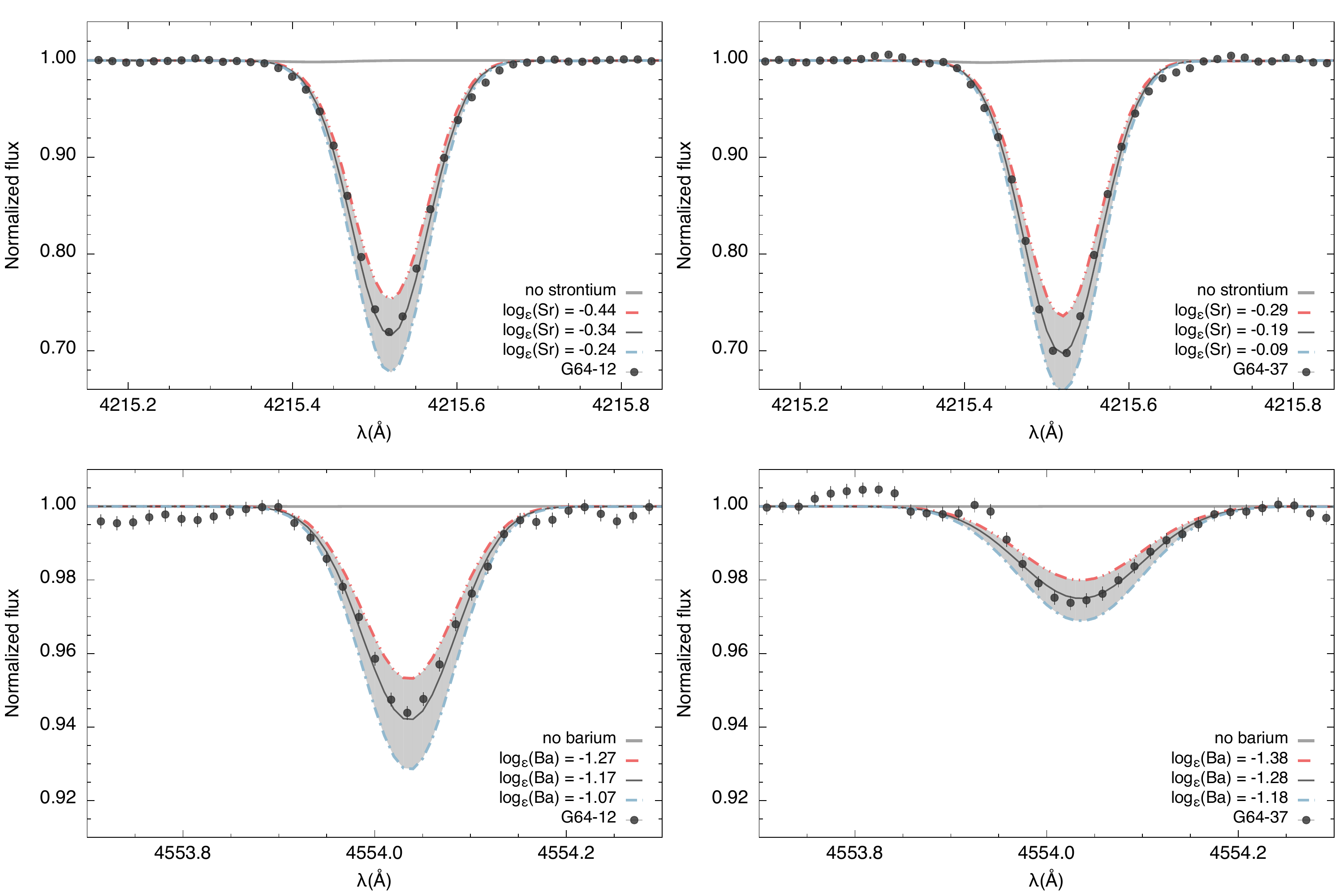}
\caption{Spectral synthesis of the \ion{Sr}{2} (upper panels) and \ion{Ba}{2}
(lower panels) features for \protect\g12\ and \protect\g37. The dots represent
the observed spectra, the solid black lines are the best abundance fits, and
the dotted and dashed line are the lower and upper abundances, used to estimate
the uncertainty.  The shaded area encompasses a 0.2~dex difference in the
abundances.  The light gray line shows the synthesized spectrum in the absence
of strontium and barium.} 
\label{basr}
\end{figure*}

In order to confirm the CEMP sub-classes of \g12\ and \g37, the carbon and barium
abundance ratios must be calculated via spectral synthesis.
The linelist used for the carbon abundance determinations has CH data from
\citet{masseron2014}, with a $^{12}$C/$^{13}$C isotopic ratio of 4 to 5
indicated therein, and atomic data from the Kurucz
linelists\footnote{\href{http://kurucz.harvard.edu/linelists.html}{http://kurucz.harvard.edu/linelists.html}}.
The final values are \cfe\ = $+1.07$ and \xfe{Ba}\ = $-0.07$ for \g12\ and \cfe\
= $+1.12$ and \xfe{Ba}\ = $-0.36$ for \g37, satisfying the original
definition for CEMP-no stars ([C/Fe] $> +1.0$, [Ba/Fe] $< 0.0$;
\citealt{beers2005}). 
The atmospheric parameters errors have an effect of $\sim
0.02$~dex on $A$(C), which does not affect the conclusions regarding the
classification of these stars.

Figures~\ref{csyn} and \ref{basr} show, respectively, the spectral
synthesis of the CH $G$-band molecular features and the \ion{Sr}{2}
($\lambda$4215) and \ion{Ba}{2} ($\lambda$4554) atomic features, for
\g12\ and \g37. The dots represent the observed spectra, the solid black
lines are the best abundance fits, and the dotted and dashed lines are the
lower and upper abundance limits, used to estimate the uncertainties in these
determinations. The shaded area encompasses a 0.2-0.4~dex difference in
the abundances. The light gray lines show the synthesized spectra in
the absence of C, Sr, and Ba. 

Both stars are also classified as CEMP-no stars by the
$A$(C) criterion described in \citet{yoon2016}. Their values of $A$(C) =
6.21 (\g12) and $A$(C) = 6.44 (\g37), with metallicities of [Fe/H] =
$-3.29$ and $-3.11$, respectively, place them among the CEMP-no Group-II
stars in Figure 1 of Yoon et al. (the Yoon-Beers diagram), although
\g37\ lies close to the CEMP-no Group-III region. However, based on
their Na and Mg abundances, both these objects are clearly Group-II
stars (see Figure 4 of Yoon et al.).

We note that the carbon abudances were determined from a 1D LTE analysis, and
non-LTE and/or 3D effects may affect the determination from the CH
absorption features. From \citet{bonifacio2009},
a star with similar parameters as \g12\ and \g37\ would have its carbon
abundance decreased by $\sim0.6$~dex if 3D effects were taken into account.
However, given that the Fe abundance is also decreased by $\sim0.3$~dex, the net
effect on \cfe\ would be $-$0.30, and both \g12\ and \g37\ would still meet 
the \cfe$\geq+0.7$ CEMP criterion of \citet{aoki2007}.

\section{Implications on First-Star Nucleosynthesis}
\label{disc}

The chemical-abundance pattern of CEMP-no Group-II stars with \metal~$<-3.0$ is
likely to be the result of one, or at most a few, supernova explosions that
occured in the early Universe \citep{ito2013,placco2014b,placco2015}, 
polluting the ISM of the natal clouds from which second-generation stars formed.
To quantify how the previously under-estimated carbon abundances affect the
inferred progenitor masses of \g12\ and \g37, we employed the
\texttt{starfit}\footnote{\href{http://starfit.org}{http://starfit.org}} online
routine, in order to compare the observed elemental abundances with an
extensive grid of non-rotating massive-star models of primordial composition
from \citet{heger2010}.

\begin{figure*}[!ht]
\epsscale{1.15}
\plotone{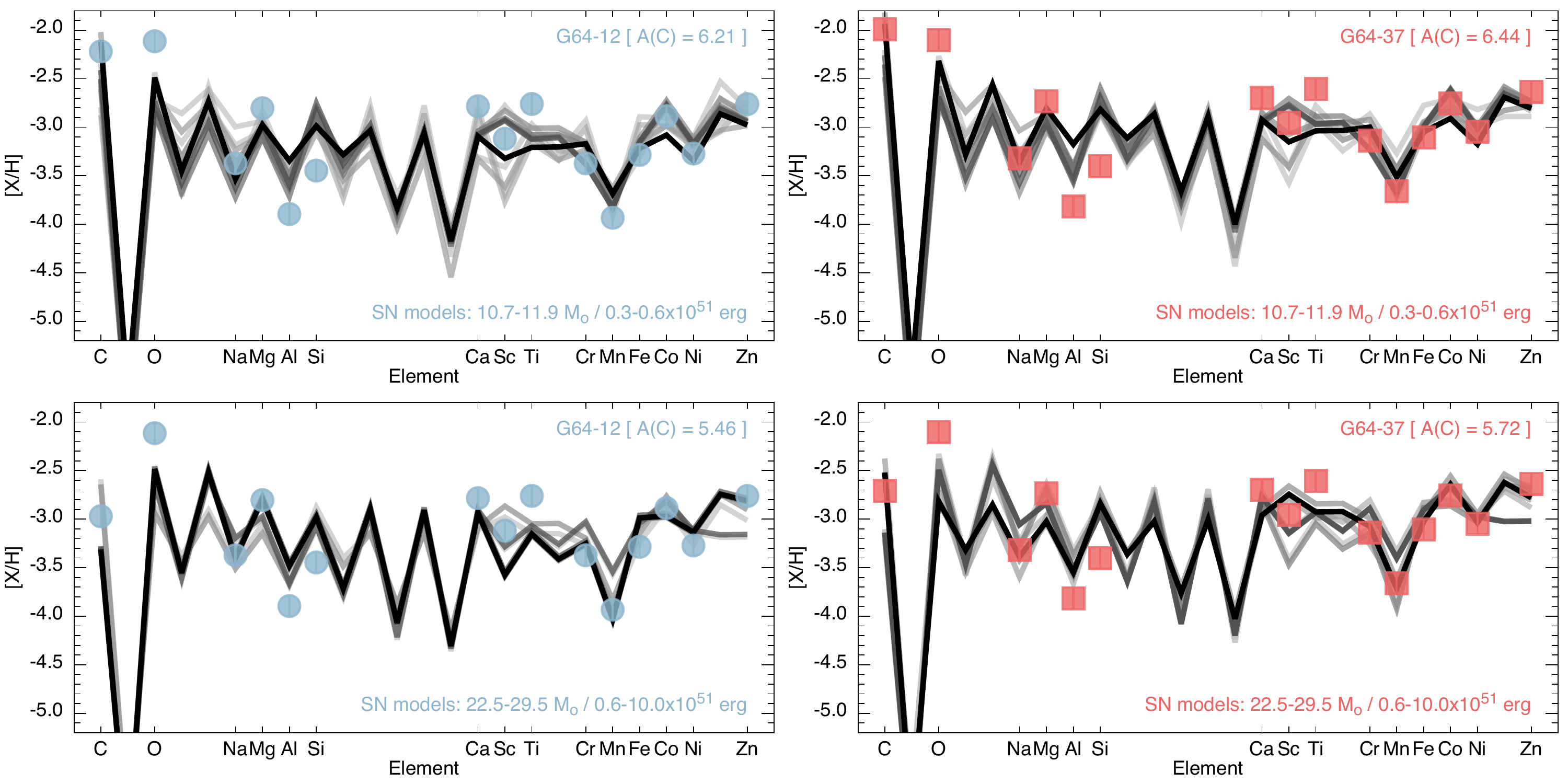}
\caption{Best model fits for \protect\g12\ (left panels) and \protect\g37\
(right panels), using the carbon abundances measured by \citet{reggiani2016}
(top panels) and the ones from the literature (lower panels). The filled symbols
represent the observed abundances in Table~\ref{abfinal}, and the solid lines
show the yields for the 10 best models fits in each case, with the mass and
explosion energy ranges listed in the lower right part of each panel.}
\label{star}
\end{figure*}
For the measured abundances in \g12, the best model-fit is a progenitor with
$10.9\,\Msun$, explosion energy of $0.6\times10^{51}\,\erg$, and a remnant mass
of $1.5\,\Msun$. If the lowest previous carbon-abundance measurement for \g12\
is used as an input \citep[$A$(C) = 5.46\footnote{The quoted
value in \citet{barklem2005} is $A$(C) = 5.66, with a measured temperature of
\teff = 6141~K. To roughly account for the $\sim400$~K difference in
temperature with respect to the present work, we re-derived the carbon abundance
using a model atmosphere with the stellar parameters from \citeauthor{barklem2005}, 
and the measured value decreased by 0.2~dex. We then applied this offset to the
quoted value, yielding $A$(C)=5.46.}
;][]{barklem2005}, the best model-fit would be a progenitor with
$29.5\,\Msun$, explosion energy of $10\times10^{51}\,\erg$, and a remnant mass
of $1.7\,\Msun$. 
For \g37, the best model-fits for the measured abundances is a
progenitor with $10.9\,\Msun$, explosion energy of $0.6\times10^{51}\,
\erg$, and a remnant mass of $1.4\,\Msun$. Using the carbon abundance
measured by \citet[][$A$(C) = 5.72, and roughly the same
\teff]{akerman2004}, the progenitor is best described by the model with
$22.5\,\Msun$, explosion energy of $10\times10^{51}\,\erg$, and a
remnant mass of $1.6\,\Msun$. 

Figure~\ref{star} shows the result of the model-fitting exercise, following the
abundance sampling procedure described in \citet{roederer2016} and
\citet{placco2016}. The left panels show the results for \g12, and the right
panels for \g37. The upper and lower panels show the fits for the different
carbon abundances listed above.  The measured abundances are represented by the
filled symbols, and the solid lines represent the yields for the 10 best 
model-fits (i.e., lowest residuals between model and observations), with the mass and
explosion energy ranges listed in the lower right part of each panel. 

From Figure~\ref{star}, it can be seen that the overall agreement is
similar for the different carbon abundances, with the exception of Al and Si, which
are over-produced by the models in the case of higher carbon abundances.  Thus,
the carbon-abundance measurements for both of these stars impacts estimates of
their progenitor masses, by factors of $\sim 2-3$ (lower for lower $A$(C)). The
effect on the predicted explosion energies is greater still, on the order of
$\sim 10-15$ (higher for lower $A$(C)), sufficient to affect interpretation of
the nature of the progenitors. Our new estimates are consistent with ``faint''
mixing-and-fallback supernovae \cite[e.g.;][and references
therein]{heger2010,nomoto2013}.

There are also implications for derivation of the frequency of
CEMP-no stars as a function of metallicity, which provides 
an important constraint for Galactic chemical evolution models
\citep{lee2013,placco2014c}.  Stars with effective temperatures near the
main-sequence turnoff, such as \g12\ and \g37, which are numerous in
extant large samples of CEMP stars such as those from the Sloan Digital
Sky Survey \citep{york2000}, could easily be
mis-classified as carbon-normal stars, resulting in a spurious reduction in
their derived frequencies.  This is particularly the case for frequencies based
on low- to medium-resolution spectroscopy of limited $S/N$ ratios.  
\citet{lee2013} quantified this effect, and found that for
stars with \metal$<-2.5$, the fraction of CEMP stars was 32\% for their giant
sample, 10\% for turnoff stars, and 15\% for main-sequence stars. This effect is
attributed to difficulties in measuring carbon abundances at high temperatures,
since the average carbon abundance for a given population of stars should not
depend on their evolutionary status, once proper corrections are taken into
account \citep[see][for further details]{placco2014c}.

For samples with carbon abundances measured in high-resolution, there is an
opposite trend. From the SAGA Database \citep{saga2008}, using stars with
\metal$< -3.0$ and carbon measurements (excluding upper limits),
there are 202 stars with \teff$< 5750$~K and 58 stars with \teff$\geq 5750$~K.
For the latter \teff\ range, 52\% of the stars are CEMP, in contrast with the overall fraction of 43\%
given by \citet{placco2014c}. This is yet another consequence of the difficulty
of measuring carbon at high temperatures, which in this case introduces a bias
towards higher values, which can still be measured at lower $S/N$ ratios.
In addition, the percentage of reported upper limits in carbon abundances
increases from 5\% for stars with \metal\ $ < -3.0$ and \teff\ $ < 5750$~K, to 34\% for
stars with \teff$\geq5750$~K (most with upper limits \cfe\ $ > +1.0$). Higher $S/N$
data for such stars is clearly desirable, and will change the CEMP frequencies
currently reported.
We note, in passing, that the star SDSS~J102915$+$172927
\citep[\metal\ $ \sim -$5.0;][]{caffau2011b}, an ultra metal-poor star with \teff\
$\sim 5800$~K and only an upper limit on its carbon abundance, may fall into
this category. If a carbon enhancement is confirmed for this star, it would
increase the CEMP fractions from 80\% to 100\% for \metal$\leq-5.0$.

\section{Conclusions}
\label{final}

In this letter we have provided conclusive evidence that the well-known
and extensively studied EMP subdwarfs \g12\ and \g37\ are in fact
CEMP-no Group-II stars, based on high-resolution, extremely high
signal-to-noise data from KECK/HIRES. 
Both are members of the outer-halo population and are single 
stars, properties that have been demonstrated to be  
associated with CEMP-no stars \citep{carollo2014,hansen2016}.
Many more examples of warm CEMP-no
stars near the halo main-sequence turnoff that have been previously
claimed as carbon-normal stars are likely to be found in future
investigations. We have also shown that under-estimates of the carbon
abundances for such stars can have large affects on the inferred
progenitor masses and explosion energies of CEMP-no stars. Future
studies of the frequencies of CEMP-no stars as a function of metallicity
will need to consider this result as well; existing frequency estimates
based on samples including stars with effective temperatures $\gtrsim
5750$~K likely under-estimate the true fractions of CEMP-no stars.

\acknowledgments

We thank Johannes Andersen, Tadafumi Matsuno, and the anonymous referee,
for providing insightful comments on the manuscript.
V.M.P. and T.C.B. \ acknowledge partial support for this work from the National
Science Foundation under Grant No. PHY-1430152 (JINA Center for the Evolution
of the Elements).
H.R. acknowledges support from the CAPES fellowship program.
J.M. acknowledges support from FAPESP (2012/24392-2) and CNPq (Bolsa de
Produtividade).
This research has made use of NASA's Astrophysics Data System Bibliographic
Services; 
the {\texttt{arXiv}} pre-print server operated by Cornell University
(\href{http://arxiv.org/}{http://arxiv.org/}); 
the SIMBAD database hosted by the Strasbourg Astronomical Data Center;
the IRAF software
packages distributed by the National Optical Astronomy Observatories, which are
operated by AURA, under cooperative agreement with the NSF
(\href{http://iraf.noao.edu/}{http://iraf.noao.edu/}); 
the SAGA Database
\citep[Stellar Abundances for Galactic Archeology;][]{saga2008}; 
the {\texttt{R-project}} software package
(\href{https://www.r-project.org/}{https://www.r-project.org/}); 
the {\texttt{gnuplot}} command-line plotting program;
and the {\texttt{stackoverflow}} online Q\&A platform
(\href{http://stackoverflow.com/}{http://stackoverflow.com/}).

\vphantom{placeholder to add a line and move references to next column}

\vphantom{placeholder to add a line and move references to next column}

\bibliographystyle{apj}

\end{document}